\documentclass[aps,prl,prlbib,epsfig,amsmath,amssymb,twocolumn,superscriptaddress]{revtex4}
\usepackage{epsfig}
\usepackage{graphics}
\newcommand{\ua}{\uparrow}
\newcommand{\da}{\downarrow}
\newcommand \pp{{\bf p}}

\begin{document}

\title{Collisional Properties of a Polarized Fermi Gas with Resonant Interactions}

\author{G.\ M.\ Bruun}
\affiliation{Dipartimento di
Fisica, Universit\`a di Trento and CNR-INFM BEC Center, I-38050
Povo, Trento, Italy}
\affiliation{Niels Bohr Institute, University of Copenhagen, DK-2100 Copenhagen \O, Denmark}
\author{A.\ Recati}
\affiliation{Dipartimento di
Fisica, Universit\`a di Trento and CNR-INFM BEC Center, I-38050
Povo, Trento, Italy}
\author{C.\ J.\  Pethick}

\affiliation{Niels Bohr Institute, University of Copenhagen, DK-2100 Copenhagen \O, Denmark}
\affiliation{NORDITA, Roslagstullsbacken 21, 10691 Stockholm, Sweden}
\author{H.\ Smith}
\affiliation{Niels Bohr Institute, University of Copenhagen, DK-2100 Copenhagen \O, Denmark}

\author{S.\ Stringari}
\affiliation{Dipartimento di
Fisica, Universit\`a di Trento and CNR-INFM BEC Center, I-38050
Povo, Trento, Italy}

\begin{abstract}

Highly polarized mixtures of atomic Fermi gases constitute a novel Fermi liquid. We demonstrate how information on thermodynamic properties may be used to calculate quasiparticle scattering amplitudes even when the interaction is resonant and apply the results to evaluate the damping of the spin dipole mode. We estimate that under current experimental conditions, the mode would be intermediate between the hydrodynamic and collisionless limits.


\end{abstract}

\maketitle

Recent experiments with polarized atomic Fermi gases have made
possible the realization of novel quantum systems~\cite{MIT}.  The
case we shall focus on here is that of a highly polarized mixture
of two fermion species.   Because the system remains normal 
at the lowest temperatures  attained, it is a completely new normal Fermi liquid.
Since the interaction may be tuned by exploiting Feshbach resonances, it is possible 
to investigate the effects of strong correlations in a previously inaccessible regime \cite{Giorgini}. The system is particularly rich, because of the ability to vary the relative numbers of the two sorts of atom and the ratio of the atomic masses, in addition to the strength of the interaction and the temperature.
In this Letter we calculate how the resonant interaction affects the frequency and damping of dipole modes in which the two components move relative to each other. A key element in the calculation is the use of thermodynamic arguments to deduce quasiparticle scattering amplitudes when the gas is strongly interacting.  Dipole modes  have previously been studied for two different spin states of $^{40}$K at higher temperatures  in a regime in which the gas is sufficiently dilute that the scattering amplitude is simply related to the scattering length \cite{jin}.  Related issues have been investigated in the context of spin-drag phenomena in low-dimensional Fermi 
systems \cite{Weber,polini}.

We consider a homogeneous  gas of two species of fermion, which
may be either two different hyperfine states of the same atom or
two different atoms, e.g., $^6$Li and $^{40}$K. We denote the
species by the label $\sigma=\uparrow,\downarrow$, the numbers of
atoms  by $N_\uparrow$ and  $N_\downarrow$,  and their masses by
$m_\uparrow$ and $m_\downarrow$.   The interaction between an
up-atom and a down-atom is characterized by the $s$-wave
scattering length $a$. Interactions between like atoms may be
neglected because the  $s$-wave component  vanishes due to the
Pauli principle. In the case of large polarization or population
imbalance, $N_\uparrow\gg N_\downarrow$, the majority (up)
component  is essentially an ideal Fermi gas with an effective mass $m^*_\ua$ equal to the bare mass $m_\ua$ even in the unitarity
limit where $|a|\rightarrow\infty$. By contrast,   the minority
(down) component is strongly affected by the interaction with the
up-atoms. The ground state energy of a single down-atom in a sea
of up-atoms can be written as
\begin{equation}
\mu_\downarrow=- \alpha\epsilon_{{\rm F}\uparrow},
\label{alpha}
\end{equation}
$\epsilon_{{\rm
F}\uparrow}=(6\pi^2n_\uparrow)^{2/3}/2m_\ua$ being the Fermi energy of
the spin-up component and $n_\sigma$ the density of the $\sigma$
atoms (we use $\hbar=1$). The parameter $\alpha$ depends on the mass ratio $m_\da/m_\ua$ and on the variable $k_{\rm F\ua}a$, where
$k_{\rm F\ua}$ is the Fermi momentum of the up-atoms. For equal
masses, Monte Carlo calculations in the unitarity limit give
$\alpha\approx 0.6$ \cite{Lobo, Pilati, Prokofiev}, and for other
mass ratios $\alpha$ has been evaluated in the ladder
approximation~\cite{Combescot} which, for equal masses,
gives good agreement with the Monte Carlo results. It is found
that $\alpha$ is an increasing function of $m_\da/m_\ua$. The
effective mass $m^*_\da$ of a down-atom is different from the bare
mass and, for $m_\da=m_\ua$, Monte Carlo calculations in the
unitarity limit give $m^*_\da \approx m_\da$~\cite{Lobo}.
Furthermore, the ladder-approximation calculations show that for
large $|a|$ the single-particle propagator of the minority
component has a large quasiparticle peak
\cite{Massignan}.

The damping of counterflow is determined by the rate at which
momentum is transferred between the two components. Consider a
situation in which the two components are spatially uniform. We
use concepts of Fermi liquid theory to describe the effects of the
interactions. The system is considered as  an ideal gas of
majority (up) atoms mixed with a gas of minority (down) atoms
whose elementary excitations are quasiparticles with effective
mass $m_\da^*$. We take the minority component to have
a mean velocity ${\mathbf{v}}$ with respect to the majority
component giving a total momentum per unit volume
${\mathbf{P}}_\downarrow=n_\downarrow m_\da^*{\mathbf{v}}$.

We define a momentum relaxation time $\tau_P$ by the relation
\begin{equation}
\frac{d{\mathbf{P}}_\downarrow}{dt}=-\frac{{\mathbf{P}}_\downarrow}{\tau_P},
\label{taudef}
\end{equation}
and we shall calculate $\tau_P$ by assuming that both components
are in thermal equilibrium described by the distribution functions
$n_{\pp'\uparrow}=f[\beta(\epsilon_{\pp'\uparrow}-\mu_{\uparrow})]$
and
$n_{{\mathbf{p}}\downarrow}=f[\beta(\epsilon_{\pp\downarrow}-{\mathbf
{p}}\cdot{\mathbf {v}}-\mu_\downarrow)]$ with $\beta=1/kT$ and
$f(x)=1/({\rm e}^x+1)$. The single particle energies are
$\epsilon_{\pp'\ua}=p'^2/2 m_\ua$ and $\epsilon_{\pp\da}=p^2/2
m_\da^*$. The term ${\mathbf {p}}\cdot{\mathbf {v}}$ boosts
the down-atom  distribution function by a velocity ${\mathbf{v}}$.
The rate of change of the
momentum of the down-atoms due to their collisions with
up-atoms may then be written as
 \begin{gather}
\frac{d{\mathbf{P}}_\downarrow}{dt}=-2\pi\frac{|U|^2}{V^3}
\sum_{{\mathbf{p}}, {\mathbf{p}}', {\mathbf{q}}}{\mathbf{p}}\left[n_{\pp\da} n_{\pp' \ua}
(1-n_{\mathbf{p-q} \da})(1-n_{\mathbf{p'+q} \ua}) \phantom{\frac{1}{x}}\right.\nonumber\\
\left.\phantom{\frac{1}{x}}-n_{\mathbf{p-q} \da}n_{\mathbf{p'+q} \ua}(1-n_{\pp \da})(1-n_{\pp' \ua})\right]  \nonumber\\
\times\delta(\epsilon_{\pp \da}+\epsilon_{\pp' \ua}-\epsilon_{\mathbf{p-q} \da}-\epsilon_{\mathbf{p'+q} \ua}),
\label{dPdt}
\end{gather}
where $V$ is the volume of the system. The two terms in
(\ref{dPdt}) correspond to a pair of quasiparticles with momenta
${\mathbf{p}}$ and ${\mathbf{p}}'$ scattering to a pair with
momenta ${\mathbf{p-q}}$ and ${\mathbf{p'+q}}$ and the inverse
process. 

The effective interaction $U$ may be estimated from
thermodynamic arguments.  The  Landau quasiparticle interaction
averaged over the angle betwen the momenta of the two
quasiparticles may be determined from the energy as a function of
the densities of the two components,
$f^0_{\ua\da}=\partial^2E/\partial n_\ua\partial n_\da=\partial
\mu_\da/\partial n_\ua$, where $E$ is the energy density of the
system.   Since the momenta of the down-atoms are assumed to be
much less than the Fermi momentum of the up-atoms, the
quasiparticle interaction may be taken to be independent of the angle
between the quasiparticle momenta.  To estimate scattering
amplitudes in terms of Landau parameters it is generally necessary
to allow for additional processes due to screening by
particle--hole pairs \cite{bp}.  However, since we assume that
$n_\da \ll n_\ua$, these processes may be neglected, and we take
the scattering amplitude to be independent of the direction of the
momenta of the quasiparticles and equal to
\begin{equation}
U= \frac{\partial \mu_\da}{\partial n_\ua}=\frac{2\pi^2}{m_\ua k_{\rm F \ua}}\gamma,
\label{U}
\end{equation}
where, from Eq.\ (\ref{alpha}), $\gamma =-\alpha[1 +(3/2)\partial \ln\alpha/\partial \ln n_\ua]$ and $k_{\rm F \sigma}=(6\pi^2n_\sigma)^{1/3}$.
For the case of a resonant interaction, $\gamma= -\alpha$ 
and $U=-(2\alpha/3)\epsilon_{{\rm
F}\uparrow}/n_\uparrow \propto 1/k_{\rm F\ua}$. This is very
different from the effective interaction at low densities, which
is proportional to $a$.

It is convenient to rewrite the expression (\ref{dPdt}) in terms of response functions. On
introducing the quantity $\omega_{\mathbf{q}}={\mathbf{q}}\cdot{\mathbf{v}}$,
using the relation
$n_\pp(1-n_{\mathbf{p-q}})=(n_\pp -
n_{\mathbf{p-q}})/\left\{1-\exp[\beta(\epsilon_\pp-\epsilon_{\mathbf{p-q}})]\right\}$,
and taking the continuum limit we obtain
\begin{gather}
\frac{d{\mathbf{P}}_\downarrow}{dt}=-2\pi|U|^2\int\frac{d^3q}{(2\pi)^3}{\mathbf{q}}\nonumber\\
\times\int_{-\infty}^{\infty}
d\omega\frac{{\rm{Im}}\chi_\downarrow(q,\omega_{\mathbf{q}}-
\omega){\rm{Im}}\chi_\uparrow(q,\omega)}{(1-{\rm e}^{\beta(\omega-\omega_{\mathbf{q}})})(1-{\rm e}^{-\beta\omega})},
\label{dPdt3}
\end{gather}
where
\begin{equation}
{\rm{Im}}\chi_\sigma(q,\omega)=\int\frac{d^3p}{(2\pi)^3}(n_{\pp
\sigma}-n_{{\mathbf{p+q}} \sigma}) \delta(\omega+\epsilon_{\pp
\sigma}-\epsilon_{\mathbf{p+q} \sigma})
\end{equation}
is, apart from a factor of $\pi$, the imaginary part of the Lindhard
function, and   the distribution functions are now global equilibrium ones
without the boost for the down-atoms. 

Let us consider first the momentum relaxation rate for $T=0$. In
this case,  the Bose factors in (\ref{dPdt3}) result in the
condition $0\le\omega\le\omega_{\mathbf{q}}$. In the following we
discuss two important limiting regimes where simple expressions
for $\tau_P$ can be obtained.

(i) The low velocity regime, $m_\da^* v\ll k_{{\rm F}\da}$.
In this case the significant contribution to (\ref{dPdt3}) comes
from $q\le 2k_{{\rm F}\downarrow}$ with a  small energy transfer
$\omega_{\mathbf{q}}\ll k_{{\rm F}\downarrow}^2/2m^*_\da$. We can then
use ${\rm{Im}}\chi_\sigma(q,\omega)={m^*_\sigma}^2\omega/(4\pi^2q)$
and the resulting integrals in (\ref{dPdt3}) yield
 \begin{equation}
\frac{1}{\tau_P}=\frac{4\pi}{25}|\gamma|^2   \left(\frac{k_{{\rm
F}\downarrow}}{k_{{\rm F}\uparrow}}\right)^2
m_\da^* v^2 
=\frac{4\pi}{25}
\frac{1}{\tau_0}
 \left(
\frac{m_\da^* v}{k_{{\rm F}\downarrow}}\right)^2, 
 \label{lowv}
\end{equation}
where $1/\tau_0 = |\gamma|^2k_{{\rm
F}\downarrow}^4/m^*_\da k_{{\rm F}\uparrow}^2$.

(ii) The high velocity regime, $k_{{\rm F}\downarrow}\ll m_\da^*
v\ll k_{{\rm F}\uparrow}$. In this case we can again carry out the
integrations in (\ref{dPdt3}) and obtain
\begin{equation}
\frac{1}{\tau_P}=\frac{2\pi}{35}|\gamma|^2 \frac{{m^*_\da}^3 v^4}{k_{{\rm
F}\ua}^2}=
\frac{2\pi}{35}\frac{1}{\tau_0}
\left(
\frac{m_\da^* v}{k_{{\rm F}\downarrow}}\right)^4
.\label{lowkf}
\end{equation}
More generally,  the scaled relaxation time
${\tilde{\tau}}_P\equiv \tau_P/\tau_0$ depends
only on the variable $\tilde{v}=m_\da^* v/k_{{\rm F}\downarrow}$
provided $m_\da^* v \ll k_{{\rm F}\uparrow }$.

\begin{figure}
\includegraphics[width=0.8\columnwidth,height=0.5\columnwidth,clip=]{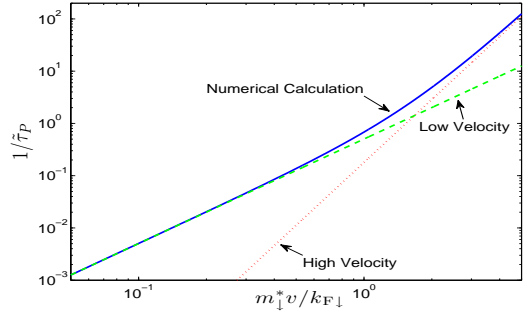}
\caption{The scaled momentum relaxation rate $1/\tilde{\tau}_P$ at $T=0$
 versus relative velocity $v$ in units of $k_{{\rm F}\da}/m^*_\da$. The full line is the result of numerical integration of Eq. (\ref{dPdt3}), the dashed line is low-velocity result  (\ref{lowv}) and the dotted
line is the high-velocity result (\ref{lowkf}).} \label{RateFig}
\end{figure}

In Fig.\ \ref{RateFig}, we plot the $T=0$ momentum relaxation rate
$1/\tilde{\tau}_P$  calculated numerically from (\ref{dPdt3}) as a function of velocity.
For the numerical calculations we took 
 $m_\da^*/m_\ua=1$ and $k_{{\rm F}\da}/k_{{\rm F}\ua}=0.1$. The relaxation rate 
increases with increasing $v$ because the available phase space for scattering grows.

We now turn to non-zero temperature.
Current experiments on highly polarized gases  achieve very low temperatures and
we therefore first analyze the regime $T\ll T_{{\rm F}\da}\ll T_{{\rm F}\ua}$, in which both
components are degenerate.  Here $kT_{{\rm F}\da} = k_{{\rm F}\da}^2/2m^*_\da$ and  $kT_{{\rm F}\ua} = k_{{\rm F}\ua}^2/2m_\ua$.  Furthermore, for small relative velocities, $  v k_{\rm F\da}\ll kT $, it is sufficient to
expand the integrand in (\ref{dPdt3}) to first order in
$\beta\omega_{\mathbf{q}}$. Using the symmetry property
${\rm{Im}}\chi_\sigma(q,\omega)=-{\rm{Im}}\chi_\sigma(q,-\omega)$
we obtain
\begin{eqnarray}
\frac{d{\mathbf{P}}_\downarrow}{dt}=-{\mathbf{v}}\frac{\pi|U|^2}{3kT}
\int\frac{d^3q}{(2\pi)^3}q^2\nonumber\\
\times\int_{-\infty}^{\infty} d\omega
\frac{{\rm{Im}}\chi_\downarrow(q,-\omega){\rm{Im}}\chi_\uparrow(q,\omega)}{(1-{\rm e}^{\beta\omega})(1-{\rm e}^{-\beta\omega})}.
\label{dPdt4}
\end{eqnarray}
For $T\ll T_{{\rm F}\da}$,  we can again use the result
${\rm{Im}}\chi_\sigma(q,\omega)={m^*_\sigma}^2\omega/(4\pi^2q)$ which
yields for the relaxation rate in the limit of low velocities the expression
\begin{equation}
\frac{1}{\tau_P}
=\frac{4\pi^3}{9}   |\gamma|^2
\frac{m^*_\da}{k_{\rm F \ua}^2}
(kT)^2
=\frac{\pi^3}{9}   \frac{1}{\tau_0}
\left( \frac{T}{T_{{\rm
F}\da}}\right)^2.
 \label{finiteT}
\end{equation}
The $T^2$-dependence is due to the fact that the 
 phase space for scattering increases with temperature.
Equation (\ref{finiteT}) shows
that for equal masses of the two components  and at unitarity  $1/\tau_P \sim kT^2/T_{F\ua}$, as one would expect on dimensional grounds because the effective interaction measured in terms of the density of states of the up-atoms is of order unity.

Next we discuss the behavior  at temperatures comparable with or
higher than $T_{{\rm F}\da}$. When the masses of the two
components are very different it becomes relevant to consider also
temperature scales characterized by
$T_0\equiv(m_{\ua}/m_{\da})T_{{\rm F}\ua}$. As an example let us
consider the case when $T_{{\rm F}\da}\ll T_0\ll T_{{\rm F}\ua}$.
In the classical regime for the minority population ($T\gg T_{{\rm
F}\da}$) we have
\begin{equation} \frac{{\rm
Im}\chi_{\downarrow}}{n_{\downarrow}}=\left(\frac{2\pi
m^*_{\downarrow}}{kTq^2}\right)^{1/2}{\rm e}^{-\omega^2m^*_{\downarrow}/2q^2kT-q^2/8m^*_{\downarrow}kT}\sinh\left(\frac{\omega}{2kT}\right).
\end{equation}
We treat two limiting cases:  a) $T_{{\rm F}\da}\ll T\ll T_0$.
Here the upper limit on the $q$-integration in (\ref{dPdt4}) may
be extended to infinity and the result of carrying out the
integrations yields again a $T^2$-dependence for $1/\tau_P$, which
differs from (\ref{finiteT}) only by the replacement of
$\pi^3/9\approx 3.45$ by $2.98$. This suggests that the
low-temperature result (\ref{finiteT}) is accurate over a much
wider temperature range. b) $T_{{\rm F}\da}\ll T_0\ll T\ll T_{{\rm
F}\da}$. Here the upper limit on the $q$-integration is $2k_{{\rm
F}\uparrow}$. Using the fact that $\omega/\sinh(\omega/2kT)$ may be
approximated by $2kT$ and $\exp(-q^2/8m^*_{\downarrow}kT)$ by 1 we
obtain, by integrating first over $\omega$ and
subsequently over $q$, the expression
${m^*_{\downarrow}}/{\tau_P}=|U|^2{m_{\uparrow}^2 k_{{\rm
F}\uparrow}^4}/{6\pi^3}$. For a dilute system with $m^*_\da \gg m_\ua$ the effective interaction is $U=2\pi
a/m_{\uparrow}$  and this result becomes $
{m^*_{\downarrow}}/{\tau_P}= k_{{\rm F}\uparrow}n_{\uparrow} \sigma$,
in agreement with the known result for the high-temperature
mobility of a heavy particle in a degenerate quantum gas, the
cross section being $\sigma=4\pi a^2$.

\paragraph*{Experimental considerations}
We now relate our results for the homogeneous case to
experimentally observable features in the presence of a trapping
potential $V_\sigma$, which will in general be different for the two species. The momentum relaxation rate is most directly probed by
exciting the spin dipole mode of a Fermi gas above the critical
polarization where the system is normal at all temperatures \cite{MIT,Lobo}.
Let us assume that the cloud of down-atoms is displaced by a distance
$\delta X$ from the equilibrium position in the harmonic trap.
Depending on the amplitude of the displacement (and consequently
on the velocity acquired by the minority component
due to the external force) as well as on the value of temperature, 
the cloud either oscillates with weak
damping around $\delta X=0$ (collisionless regime) or it 
relaxes towards equilibrium without any oscillations (hydrodynamic
regime).

 In the collisionless limit the frequency of the oscillation is readily
obtained in the case of large imbalance where it is sufficient to
consider the single quasiparticle Hamiltonian to describe the motion of the minority component \cite{longNormalTN}.  
The interaction energy of a down-atom is $-\alpha \epsilon_{\rm F \ua}$.  In the Thomas-Fermi 
approximation,  $\epsilon_{\rm F \ua}+V_\ua$ is a constant. Thus, the  total potential felt by a down-atom is 
$V_\da+\alpha V_\ua$. The Hamiltonian for a single down-atom then 
has the form $H_{\rm sp}=p^2/2m^*_\da+V_\da+\alpha V_\ua$.
The interaction with the majority component is taken into account
through the effective mass $m^*_\da$ and  the change in the
potential caused by the interaction with the up-atoms. For simplicity we restrict ourselves to a resonant interaction, in which case $\alpha$ is independent of density and from
this Hamiltonian, the frequency $\omega_{\rm D}$ of the spin dipole mode
for a harmonic trap is easily calculated to be \cite{Lobo,longNormalTN}
\begin{equation}
\omega_{\rm D}=\omega_\da \sqrt{\frac{m_\da}{m_\da^*}\left(1+\frac{m_\ua \omega_\ua^2}{m_\da \omega_\da^2}\alpha \right) },
 \end{equation}
 where $\omega_\sigma$ is the oscillation frequency in the trap for species $\sigma$.
Measurements of the spin dipole frequency thus provide a unique
opportunity to test directly the effects of interactions which,
according to the theoretical estimates of $\alpha$ and $m^*_\da$,
should increase the value of the frequency by a factor $1.23$ when the trapping potential is harmonic and the same for the two species. The spin dipole mode, however, is well defined only
in the  collisionless limit $\omega_{\rm D}\tau_P\gg 1$. It becomes
overdamped in the  hydrodynamic regime $\omega_{\rm D}\tau_P\ll 1$ since
the spin current is not conserved by collisions \cite{vichi}. 

\begin{figure}
\includegraphics[width=0.8\columnwidth,height=0.5\columnwidth]{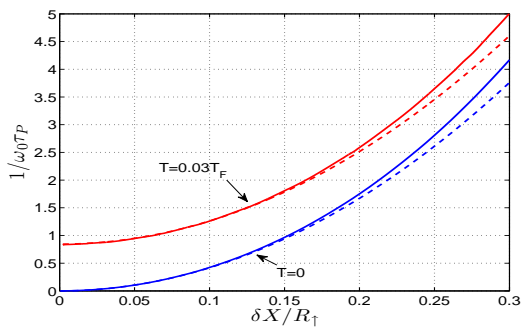}
\caption{The quantity $1/\omega_0\tau_P$ determining the damping of the dipole mode as a function of the amplitude of the oscillation for $T=0$ and $T=0.03\:T_{\rm F \ua}$. (For details see text.)}
\label{Rate2Fig}
\end{figure}

In order to estimate whether under current experimental conditions the
spin dipole mode will be in the hydrodynamic or in the
collisionless regime, we calculate $\omega_{\rm D} \tau_P$.  It is convenient to express results in terms of the amplitude of the displacement of the down-atom cloud $\delta X$, which is  controllable experimentally. 
We shall assume that the displacement of the down-atom cloud is sufficiently
small ($\delta X\ll R_\ua$  where
$R_\ua$ is the radius of the majority cloud) that  the  density of up-atoms may be regarded as uniform when estimating the relaxation rate.  
The relative velocity of the two components is given by $v=\omega_{\rm D}\delta X$.
We shall adopt the values $N_\ua=10^7$, and
$N_\da/N_\ua=0.026$ ($T_{{\rm F}\da}/T_{{\rm F}\ua}=0.3$)
corresponding to conditions achieved in the MIT experiment
\cite{MIT} for a mixture of  $^6$Li-atoms in two different
hyperfine states, together with the values $\alpha=0.6$ and
$m^*_\da/m_\da\approx1 $ obtained theoretically \cite{Combescot,Lobo}.  We approximate $\omega_{\rm D}$ by the trap frequency $\omega_0$, which we take to be the same for both species. The lower full line in Fig. \ref{Rate2Fig} shows $1/\omega_0\tau_P$ as a function of $\delta X/R_\ua$ obtained from Eq.\ (\ref{dPdt3}) by numerical integration for $T=0$,
 while the lower dashed line is the expression (\ref{lowv}).  The upper lines are for a temperature $T=0.03T_{\rm F\ua}$, the full one being the result of a numerical calculation and the dashed line is the sum of the 
 results (\ref{lowv}) and (\ref{finiteT}) which, expressed in terms of the number of up-atoms, are
\begin{gather}
\frac{1}{\omega_0 \tau_P}=
\frac{8\pi}{25}(6N_\uparrow)^{1/3}\alpha^2\frac{m_\da^*}{m_\ua}\left(
\frac{T_{{\rm F}\da}}{T_{{\rm F}\ua}}\right)^2\left( \frac{\delta X}{R_\ua}\right)^2 \label{Expv2}
\end{gather}
and
\begin{equation}
\frac{1}{\omega_0 \tau_P}=
\frac{2\pi^3}{9}(6N_\uparrow)^{1/3}\alpha^2
\frac{m_\da^*}{m_\ua}\left(\frac{T}{T_{{\rm F}\uparrow}}\right)^2,
\label{ExpT}
\end{equation}
where we have used the result $\gamma=-\alpha$ for a resonant interaction and the fact that $kT_{{\rm F}\ua} =k^2_{{\rm
F}\ua}/2m_\ua=(6N_\ua)^{1/3}\omega_0$.
The plots demonstrate that the analytical results are a good approximation to those obtained by direct numerical integration in the regimes of experimental interest. The sum of the results (\ref{lowv}) and (\ref{lowkf}), which is not shown, is an even better approximation to the numerical results.

The calculated values of $\omega_0\tau_P$ demonstrate that, for the experimental conditions now attainable at MIT,
the polarized normal phase is in a regime intermediate between
collisionless and hydrodynamic behavior, implying significant
damping of  the spin dipole mode. At lower temperature, the gas
enters the collisionless regime.

How important collisions are in a given mode is sensitive to the anisotropy of the trap, which we have neglected so far. 
For instance, for a cigar-shaped trap ($\omega_z<\omega_\perp$)
the transverse mode will be more collisionless, the value of $1/\omega_{\rm D}\tau_P$ being multiplied by a factor $(\omega_z/\omega_\perp)^{1/3}$,  for  a fixed 
value of $(\omega_\perp^2\omega_z)^{1/3}$. 
When the two species are different elements, the value of $\omega_0 \tau_P$ will be depend on the trapping potentials of the two species, which can be varied independently of each other.

For low velocity, $m_\da^* v\ll k_{{\rm F}\da}$, one sees from
(\ref{lowv}) and (\ref{finiteT}) that the momentum relaxation rate
scales as $m_\da^*$. Consequently, since $m_\da^*\approx
m_\da$  the spin motion can be made more collisionless by trapping
an atom mixture with a lighter minority component. However,
calculations indicate that this effect is reduced
due to the fact that, at unitarity, the scattering amplitude 
for the case of extreme imbalance increases
with decreasing $m_\da/m_\ua<1$~\cite{Combescot}. 
For $m_\da/m_\ua>1$  the scattering amplitude is
predicted to be approximately constant  and therefore $1/\tau_P \propto  m_\da$ in this regime. Thus,  the
spin motion becomes more hydrodynamic for $m_\da/m_\ua>1$.  It
would be interesting to test these predictions experimentally.

In conclusion, we have demonstrated that for a strongly polarized atomic gas with resonant interactions, scattering amplitudes exhibit a universal behavior, just as  thermodynamic properties do \cite{ho}.  Predictions for the damping of the spin dipole mode have been presented, and it would be valuable to make measurements of the mode. Our approach may be extended to less highly polarized gases by including the effects of screening by the minority component. We acknowledge fruitful discussions with L.\ P.\ Pitaevskii. C.\ J.\ P.\ is grateful to ECT* for 
hospitality during the initial stages of this collaboration. A.\ R.\ and S.\ S.\
  acknowledge support from the Euroquam FERMIX program.


\vspace{-2em}
\begin{thebibliography}{99}
\vspace{-1.5em}
\bibitem{MIT}
Y. Shin, M. W. Zwierlein, C. H. Schunck, A. Schirotzek, and W. Ketterle, Phys. Rev. Lett. {\bf 97}, 030401 (2006);
Y. Shin, C. H. Schunck, A. Schirotzek, and W. Ketterle, Nature \textbf{451}, 689 (2008).

\bibitem{Giorgini} For a review of atomic Fermi gases see, e.g.,
S.\ Giorgini, L.\ P.\ Pitaevskii, and S.\ Stringari, arXiv:0706.3360.

\bibitem{jin} S.\ D.\ Gensemer and D.\ S.\  Jin,  Phys.\ Rev.\ Lett.\ \textbf{87}, 173201 (2001);
B.\ DeMarco and D.\ S.\ Jin,  Phys.\ Rev.\ Lett.\ \textbf{88}, 040405 (2002).

\bibitem{Weber}C.\ P.\ Weber \textit{et al}., Nature \textbf{437}, 1330 (2005).

\bibitem{polini} See, e.g., M.\ Polini and G.\ Vignale,  Phys. Rev. Lett. \textbf{98}, 266403 (2007) and references therein.

\bibitem{Lobo}
C. Lobo,  A. Recati, S. Giorgini, and S. Stringari, Phys. Rev. Lett. \textbf{97}, 200403 (2006).

\bibitem{Pilati}
S.\ Pilati and  S.\ Giorgini,  arXiv:0710.1549.

\bibitem{Prokofiev}
N. Prokof'ev and B. Svistunov, Phys. Rev. B {\bf 77}, 020408(R) (2008).


\bibitem{Combescot}
R.\ Combescot, A.\ Recati, C.\ Lobo, and F.\ Chevy, Phys.\ Rev.\ Lett.\ \textbf{98}, 180402 (2007).

\bibitem{Massignan}
P.\ Massignan, G.\ M.\ Bruun, and H.\ T.\ C.\ Stoof,  Phys.\ Rev.\ A {\bf 77}, 031601(R) (2008).


\bibitem{bp}G.\ Baym and C.\ J.\ Pethick, {\it Landau Fermi-liquid Theory: Concepts and Applications}
(Wiley, New York, 1991).


\bibitem{longNormalTN}
A. Recati, C. Lobo, and S. Stringari,  arXiv:0803.4419.


\bibitem{vichi} L.\ Vichi and S.\ Stringari,  Phys.\ Rev.\ A {\bf 60}, 4734 (1999).

\bibitem{ho} T. L. Ho, Phys.\ Rev.\ Lett.\ \textbf{92}, 090402 (2004).

\end{thebibliography}
 \end{document}